\documentclass[twocolumn,english,showpacs,amsmath,pra]{revtex4}
\usepackage[T1]{fontenc}
\usepackage[latin9]{inputenc}
\usepackage{float}
\usepackage{amsmath}
\usepackage{graphicx}
\usepackage{amssymb}

\makeatletter

\providecommand{\tabularnewline}{\\}

\@ifundefined{textcolor}{}
{%
 \definecolor{BLACK}{gray}{0}
 \definecolor{WHITE}{gray}{1}
 \definecolor{RED}{rgb}{1,0,0}
 \definecolor{GREEN}{rgb}{0,1,0}
 \definecolor{BLUE}{rgb}{0,0,1}
 \definecolor{CYAN}{cmyk}{1,0,0,0}
 \definecolor{MAGENTA}{cmyk}{0,1,0,0}
 \definecolor{YELLOW}{cmyk}{0,0,1,0}
 }

\@ifundefined{showcaptionsetup}{}{%
 \PassOptionsToPackage{caption=false}{subfig}}
\usepackage{subfig}
\makeatother

\usepackage{babel}

\begin{document}

\title{On the Nature of Optical Excitations in Hydrogenated Aluminium Cluster
Al$_{\text{4}}$H$_{\text{6}}$: A Theoretical Study }

\author{Sridhar Sahu and Alok Shukla}

\address{Department of Physics, Indian Institute of Technology, Bombay, Powai,
Mumbai 400076, INDIA}

\email{sridhar@phy.iitb.ac.in, shukla@phy.iitb.ac.in}
\begin{abstract}
In this paper, we present a theoretical investigation of photoabsorption
spectrum of the newly synthesized hydrogenated cluster of aluminium,
Al$_{\text{4}}$H$_{\text{6}}$. The calculations are performed within
the wave-function-based semi-empirical method employing the complete
neglect of differential overlap (CNDO) model, employing a large-scale
configuration interaction (CI) methodology, and our results are found
to be in very good agreement with the earlier ones obtained from the
\emph{ab initio} time-dependent density functional theory (TDDFT).
We carefully analyze the many-particle wave functions of various excited
states up to 8 eV, and find that they are dominated by single-particle
band to band excitations. This is in sharp contrast to bare aluminium
clusters, in general, and Al$_{\text{4}}$, in particular, whose optical
excitations are plasmonic in nature. We attribute this difference
to be a consequence of hydrogenation.
\end{abstract}


\maketitle

\section{Introduction\label{sec:Intro-alh}}

Because of the future importance of hydrogen as a fuel, now-a-days
a considerable amount of research activity is taking place in the
field of hydrogen storage materials\cite{fuel-energy,crabetree}.
This is because hydrogen is an extremely combustible gas, therefore,
the preferred way to store it is in physically or chemically adsorbed
forms. It is from this point of view that a large amount of theoretical
and experimental research is taking place in the field of metallic
hydrides. Among all possibilities, hydrides of group III elements,
boron and aluminium, which are called boranes and alanes, respectively,
are considered to be strong candidates for the purpose\cite{ozturk,Zidan,aldridge}.
It is with this aim in mind, that Li \emph{et al}.\cite{XLi} recently
synthesized and studied the properties of a new alane, Al$_{\text{4}}$H$_{\text{6}}$.
The same group followed up this work, by performing an experimental
and theoretical study of the electronic structure and properties of
closo-alanes, Al$_{\text{4}}$H$_{\text{4}}$, Al$_{n}$H$_{n+2}$,
$4\leq n\leq8$\cite{alh-grubisic}. Besides their possible applications,
alanes are of interest from a fundamental point of view as well. Boron
and aluminium are in the same group of periodic table with three valence
electrons each, yet their properties are very different. In the bulk
form boron is a semiconductor, while aluminium is a metal. As far
as the hydride chemistry is concerned, boron exhibits a huge variety
of boranes with n-vertex polyhedral structures\cite{McKee,schleyer},
while the number and types of alanes is much smaller. The pioneering
works of Lipscomb\cite{lipscomb} which provide the explanations of
the cage structures of boranes using the concepts of three-centered
two-electron(3c-2e) bonding, ultimately resulted in an ingenious molecular
orbital theory known as polyhedral skeletal electron pair theory (PSEPT)
or simply Wade-Mingos rule\cite{wade,mingo,wade2}. On the other hand
only a small number of hydrides of aluminium have been investigated
theoretically and experimentally such as molecules AlH\cite{AlH-pelissier,AlH-Kaur,alh-szanja,alh-langhoff,alh-ex-marinell},
AlH$_{\text{3}}$\cite{AlnHm-early}, Al$_{\text{2}}$H$_{\text{4}}$\cite{al2h4-al2h6-wang}
Al$_{\text{2}}$H$_{\text{6}}$\cite{AlnHm-early,b-al-ga-hydrides-schaefer,b2h6-al2h6-ga2h6-barone,al2h4-al2h6-wang,al2h6-mol-vibr-andrews,al2h6-prl-rao-jena},
of the general type Al$_{n}$H$_{n+2}$\cite{AlnHn+2-martinez,Fu-AlnHn+2-alane-borane},
and cage-like clusters such as Al$_{13}$H$^{-}{}_{m}$\cite{al13hm-ion-charkin}.
Fu \emph{et al.}\cite{Fu-AlnHn+2-alane-borane} in a recent comparative
study investigated theoretically Al$_{n}$H$_{n+2}$ clusters, and
their borane analogues. Grubisic \emph{et al}.\cite{alh-grubisic}
have tried to arrive at a Wade-Mingos type of rule set for aluminium
hydrides, so that their structures could be predicted in a way similar
to boranes, though Martinez and Alonso\cite{AlnHn+2-martinez} have
recently raised doubts on this similarity. With this background in
mind, the recent study of Al$_{\text{4}}$H$_{\text{6}}$,\cite{XLi}
a material whose structure is consistent with Wade-Mingos rule, has
again revived the analogy of aluminium hydrides with boranes. In this
work we present an extensive study of the electronic structure and
optical properties of Al$_{\text{4}}$H$_{\text{6}}$, with the aim
of predicting its photoabsorption spectrum, and to understand the
nature of its optical excitations, which can be used for optical characterization
of this substance. For the purpose, we use our recently developed
CNDO/INDO Hamiltonian-based semiempirical multi-reference singles-doubles
configuration interaction (MRSDCI) methodology described elsewhere\cite{shukla,sahu-b12,boroz-sahu}.
To benchmark our approach, we also apply this approach to AlH molecule
and demonstrate good agreement with the experimental results. Martinez
and Alonso\cite{AlnHn+2-martinez} calculated the photoabsorption
spectra of several alanes of the type Al$_{n}$H$_{n+2}$ including
Al$_{\text{4}}$H$_{\text{6}}$, using the \emph{ab initio} time-dependent
density functional theory (TDDFT), and our results are found to be
in very good agreement with their results\cite{AlnHn+2-martinez}.
Upon analyzing the many-particle wave functions of the excited states
corresponding to important peaks in the spectrum of Al$_{\text{4}}$H$_{\text{6}}$,
we conclude that all the way up to 8 eV the states correspond to inter-band
excitations, and not to plasmonic excitations common in bare Al clusters.

The remainder of this paper is organized as follows. In section \ref{sec:Theory-al4h6}
we briefly describe the theoretical methodology employed for the present
calculations. This is followed by the presentation and discussion
of our results in section \ref{sec:Results-alh}. Finally, in section
\ref{sec:Conclusions-al4h6} we present our conclusions.

\section{Theory\label{sec:Theory-al4h6}}

For our study, we adopted a wave-function based electron-correlated
approach employing the semi-empirical valence-electron CNDO/2 model
Hamiltonian developed by Pople and coworkers\cite{CNDO-1a,CNDO-1b,CNDO-2}.
The methodology adopted in this work is discussed in detail in our
earlier papers\cite{shukla,sahu-b12,boroz-sahu}, therefore, we present
only a brief description of it here. As compared to our earlier works
on boron-based clusters\cite{sahu-b12,boroz-sahu} where we had used
the INDO model, the choice of CNDO/2 model here has to do with the
fact that INDO parameters are not available for the third row atoms
like aluminium in the original Pople approach\cite{INDO}. Our calculations
are initiated at the Hartree-Fock (HF) level, within the CNDO/2 model,
using a computer program developed recently by us\cite{shukla}. The
CNDO-HF molecular orbitals (MOs) thus obtained, are used to transform
the Hamiltonian from the original atomic-orbital (AO) to the MO representation,
which is subsequently used in the post-HF correlated calculations.
The transformed CNDO/2 Hamiltonian matrix elements in the MO representation
are supplied to the computer program package MELD\cite{meld}, which
is used to perform the correlated calculations using the multi-reference
singles-doubles configuration-interaction (MRSDCI) approach. Using
the ground- and excited-state wave functions obtained from the MRSDCI
calculations, electric dipole matrix elements are computed and subsequently
utilized to compute the linear absorption spectrum, under the electric-dipole
approximation, assuming a Lorentzian line shape. By analyzing the
excited states contributing to the peaks of the computed spectrum
obtained from a given calculation, bigger MRSDCI calculations are
performed with a larger number of reference states. This procedure
is repeated until the computed spectrum converges within an acceptable
tolerance. In the past, we have used such an iterative MRSDCI approach
on a number of conjugated polymers to perform large-scale correlated
calculations of their linear and nonlinear optical spectra\cite{mrsd-calc}.

\section{Results and Discussion\label{sec:Results-alh}}

In this section we present and discuss our results on the electronic
structure and optical properties computed using our CNDO-MRSDCI approach.
However, before that, to benchmark our methodology we present and
discuss our CNDO-CI results on the simplest hydride of aluminium,
namely AlH.

\subsection{AlH Molecule\label{sub:AlH-Molecule}}

First we performed the geometry optimization of the AlH molecule at
the CNDO-HF level, by calculating the total energy of the system for
different bond lengths, and then locating the minimum. Our optimized
bond length was found to be 1.75 \AA\  which is 0.1 \AA\  larger
than the reported experimental value of 1.65 \AA\cite{AlH-pelissier,AlH-Kaur}.
However, our aim here is to compute the optically-active excited states
of the molecule, and as we will show later that this much of difference
in the geometry leads to insignificant differences in their excitation
energies. 

In the CNDO/2 model Al has nine Slater-type basis functions (1s, 3p,
5d), while H has only one basis function, leading to ten basis functions
in all. Being a valence electron approach, the total number of electrons
being explicitly considered for AlH within the CNDO/2 model is four,
with Al contributing three electrons, and H one. Therefore, with four
electrons, and ten basis functions full-CI (FCI) calculations for
the system are feasible, because in the singlet subspace the total
number of all possible configuration state functions (CSFs) is only
825. Thus, the FCI results presented here are exact within the chosen
model (CNDO/2), and any disagreements with the experiments will indicate
the deficiency of the model rather than that of the CI expansion. 

The CNDO-HF calculations predict the Mulliken population of Al(H)
atoms to be +0.24(-0.24), implying a partial ionic character to this
system, with Al being the electron donor, fully consistent with previous
works\cite{al2h6-prl-rao-jena}. The absorption spectrum of AlH is
well-known and has been studied extensively over the years both experimentally\cite{alh-szanja}
and theoretically\cite{AlH-Kaur,alh-ex-marinell,alh-langhoff}. The
ground state of the system is classified as $X^{1}\Sigma^{+}$ while
the dipole connected excited states are $A^{1}\Pi$, $C^{1}\Sigma^{+}$,
$D^{1}\Sigma^{+}$, and $E^{1}\Pi$. The first excited state $A^{1}\Pi$
corresponds to HOMO($H$)$\rightarrow$LUMO($L$) transition which
is $5\sigma\rightarrow2\pi$ in nature. The reported experimental
value of this transition is 2.91 eV, with which our calculations are
in good agreement with the computed values of 3.11 eV, and 3.12 eV
at the bond lengths of 1.75 \AA\  (our optimized bond length) and
1.65 \AA\ (experimental bond length) respectively. This also shows
that small differences in geometry have negligible impact on the calculated
excitation energy of the lowest excited state. The $C$ and $D$ $\Sigma$-bands
correspond to Rydberg-type excitations and, therefore, cannot be computed
using CNDO approach, because they require the use of extremely diffused
basis functions\cite{AlH-Kaur}. The $E^{1}\Pi$ band corresponds
to $|H-1\rightarrow L\rangle$ ($4\sigma\rightarrow2\pi$) excitation,
and has been measured close to 6.9 eV\cite{alh-szanja}, while our
calculation predicts its value to be 7.1 eV, thus overestimating it
by 0.2 eV. We note that quantitatively speaking, the disagreement
between our results, and the experimental ones, both for the $A^{1}\Pi$,
and $E^{1}\Pi$ bands is 0.2 eV, which is quite satisfying, given
the semiempirical nature of this approach. Having benchmarked the
CNDO-CI approach for the case of the AlH molecule, next we present
the results of our calculations for Al$_{\text{4}}$H$_{\text{6}}$.

\subsection{Al$_{\text{4}}$H$_{\text{6}}$\label{sec:al4h6}}

\begin{figure}[H]
\includegraphics[width=6.5cm]{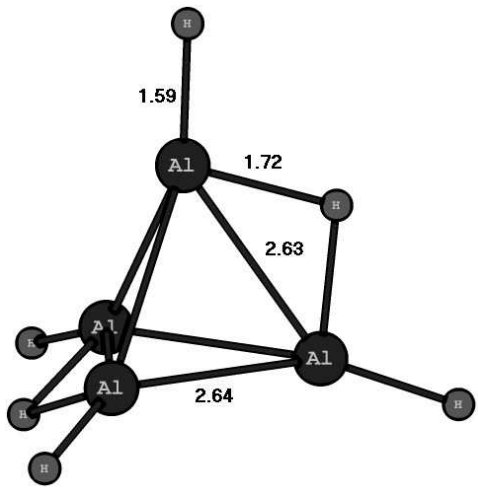}

\caption{Optimized geometry of Al$_{\text{4}}$H$_{\text{6}}$ obtained from
the \emph{ab initio} DFT calculations (see text for details). Each
bond length (in \AA) has also been indicated.}

\label{Fig-struct-al4h6}
\end{figure}

We first carried out the geometry optimization of Al$_{\text{4}}$H$_{\text{6}}$
within a density functional theory (DFT) based approach, using Gaussian03
computational package\cite{gaussian03} employing B3LYP functional,
and 6-311+g(d) basis set. The optimized geometry is presented in Fig.\ref{Fig-struct-al4h6},
and it is virtually identical to that reported by Li \emph{et al}.\cite{XLi}
As noted by Li \emph{et al}.\cite{XLi} also, this structure is very
similar to the computed structure of the borane, B$_{\text{4}}$H$_{\text{6}}$\cite{b4h6-neu}
which still has not been synthesized. We used this DFT optimized geometry
to perform the calculations of the photoabsorption spectrum within
our CNDO-CI approach, as well as the \emph{ab initio} TDDFT method. 

. 

\begin{figure}
\subfloat[HOMO-3]{\includegraphics[width=2.5cm]{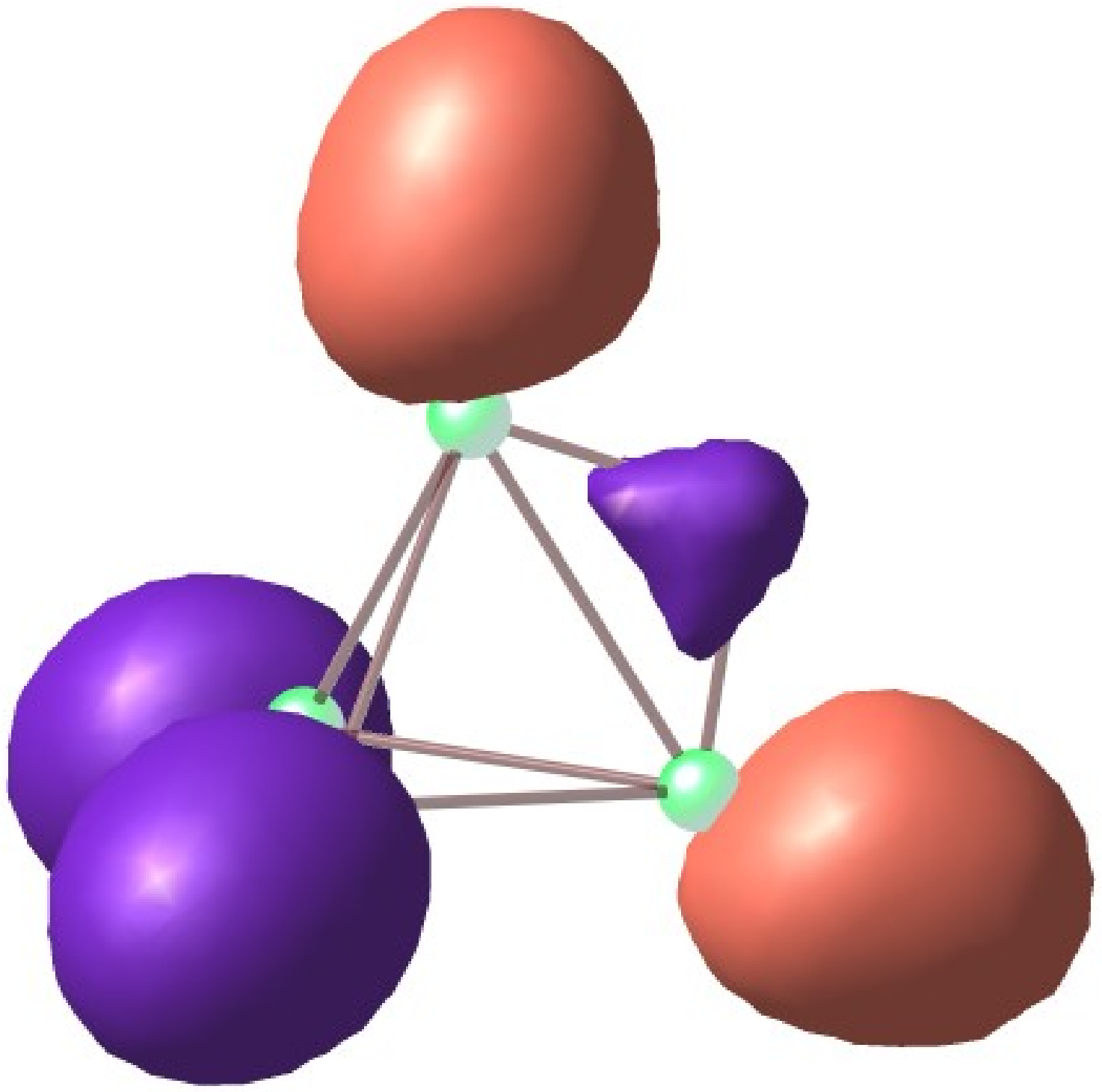}

}\subfloat[HOMO-2]{\includegraphics[width=2.5cm]{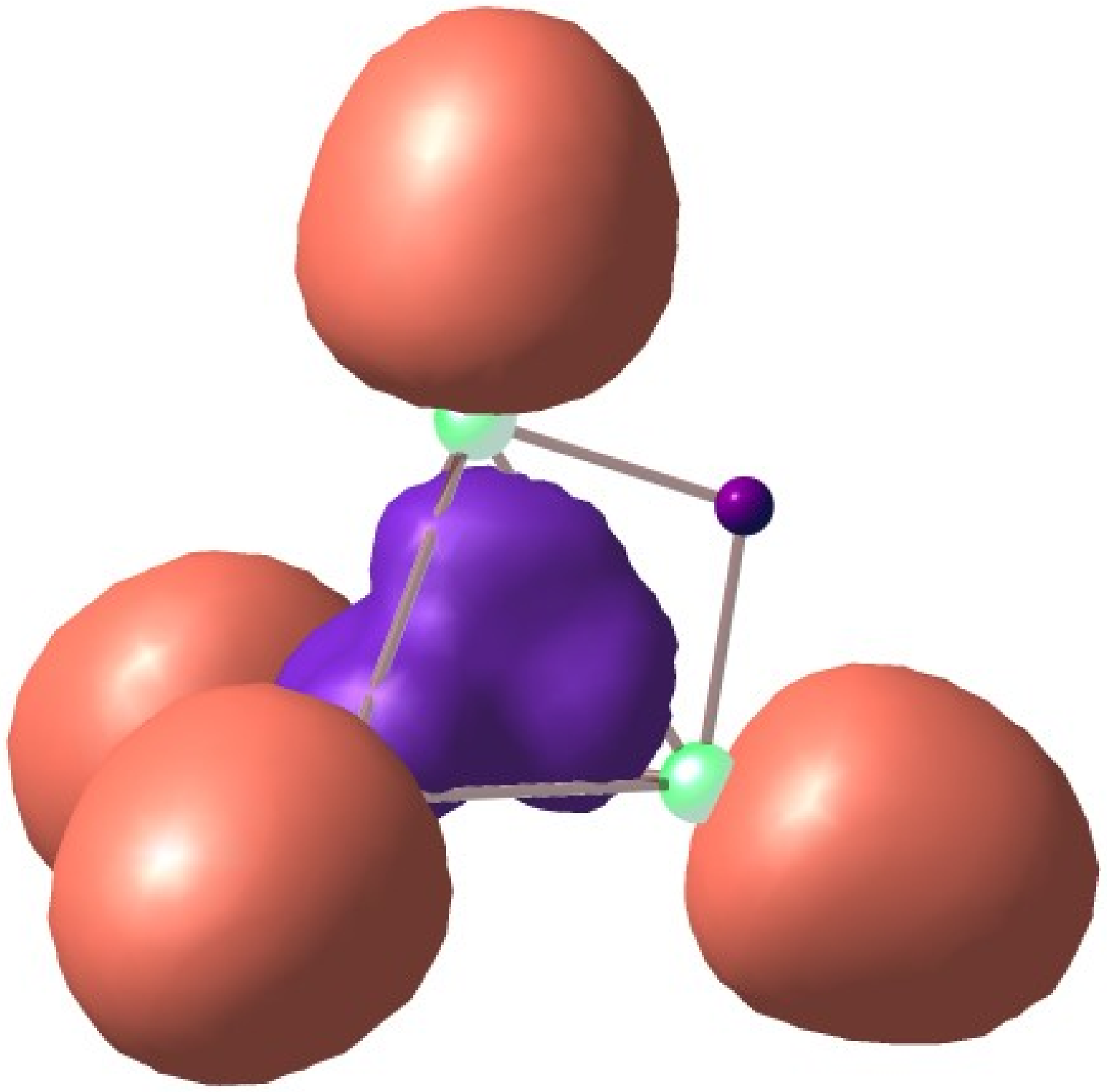}

}\subfloat[HOMO-1]{\includegraphics[width=2.5cm]{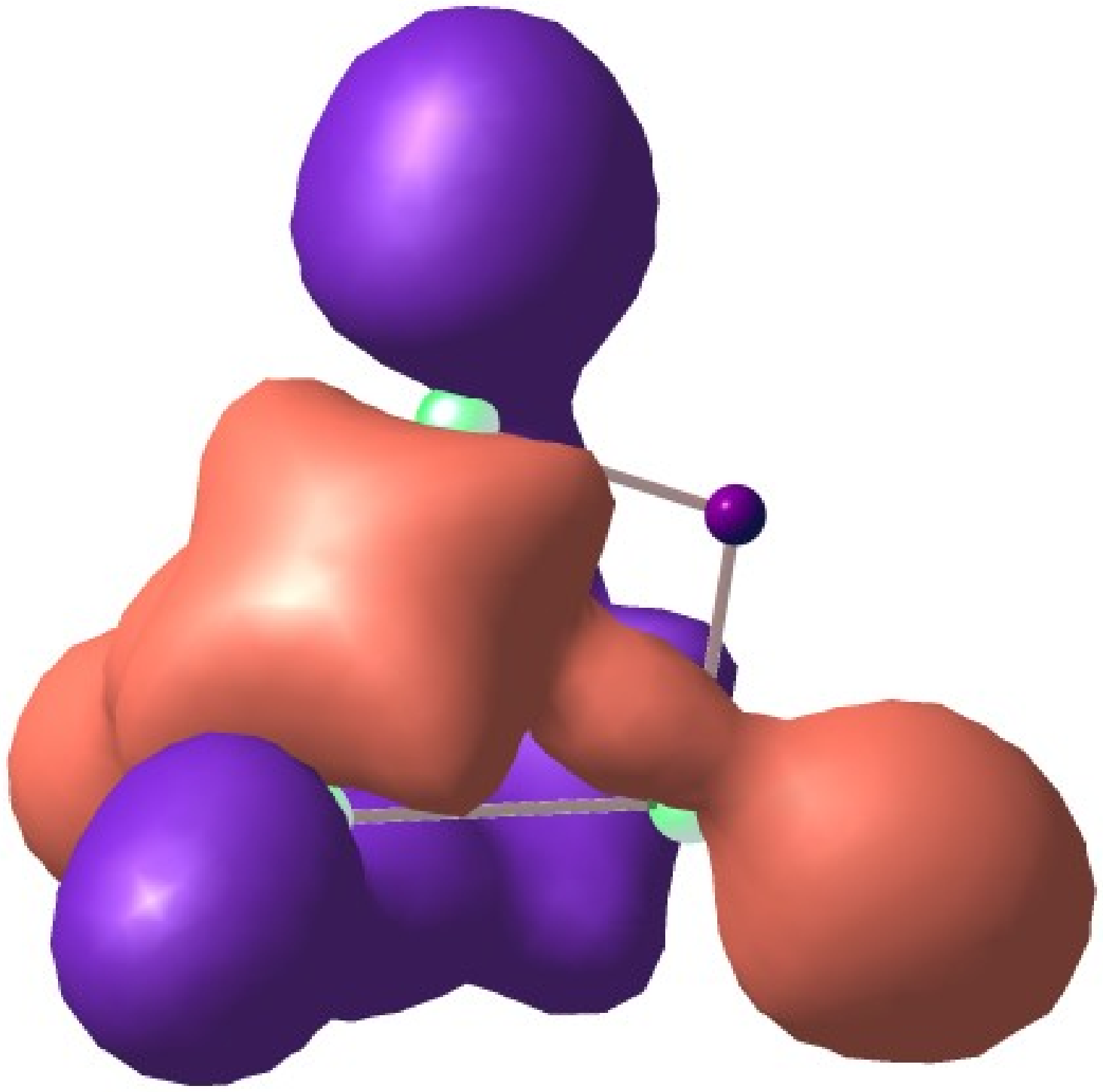}

}

\subfloat[HOMO]{\includegraphics[width=2.25cm]{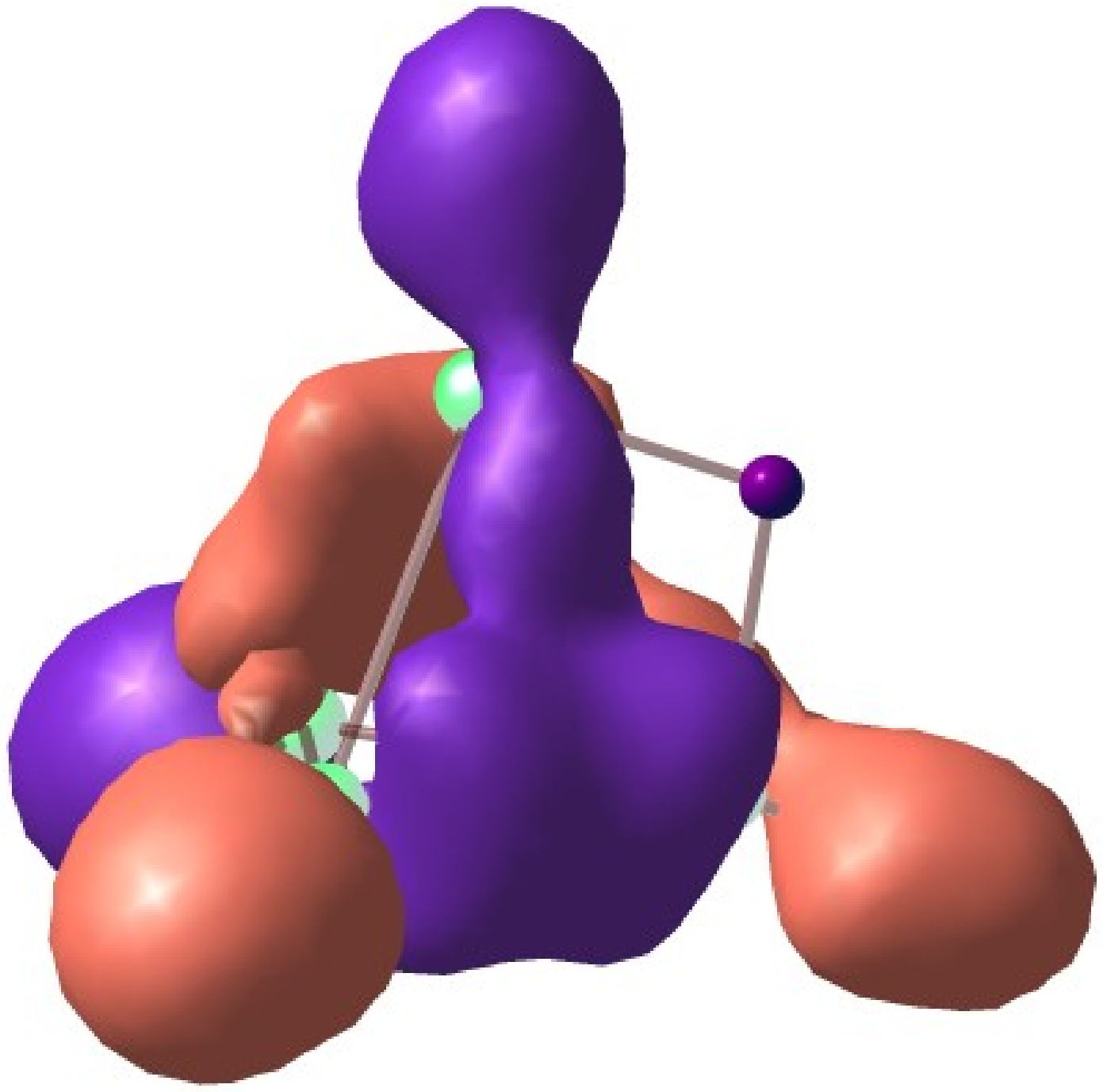}

}\subfloat[LUMO]{\includegraphics[width=2.25cm]{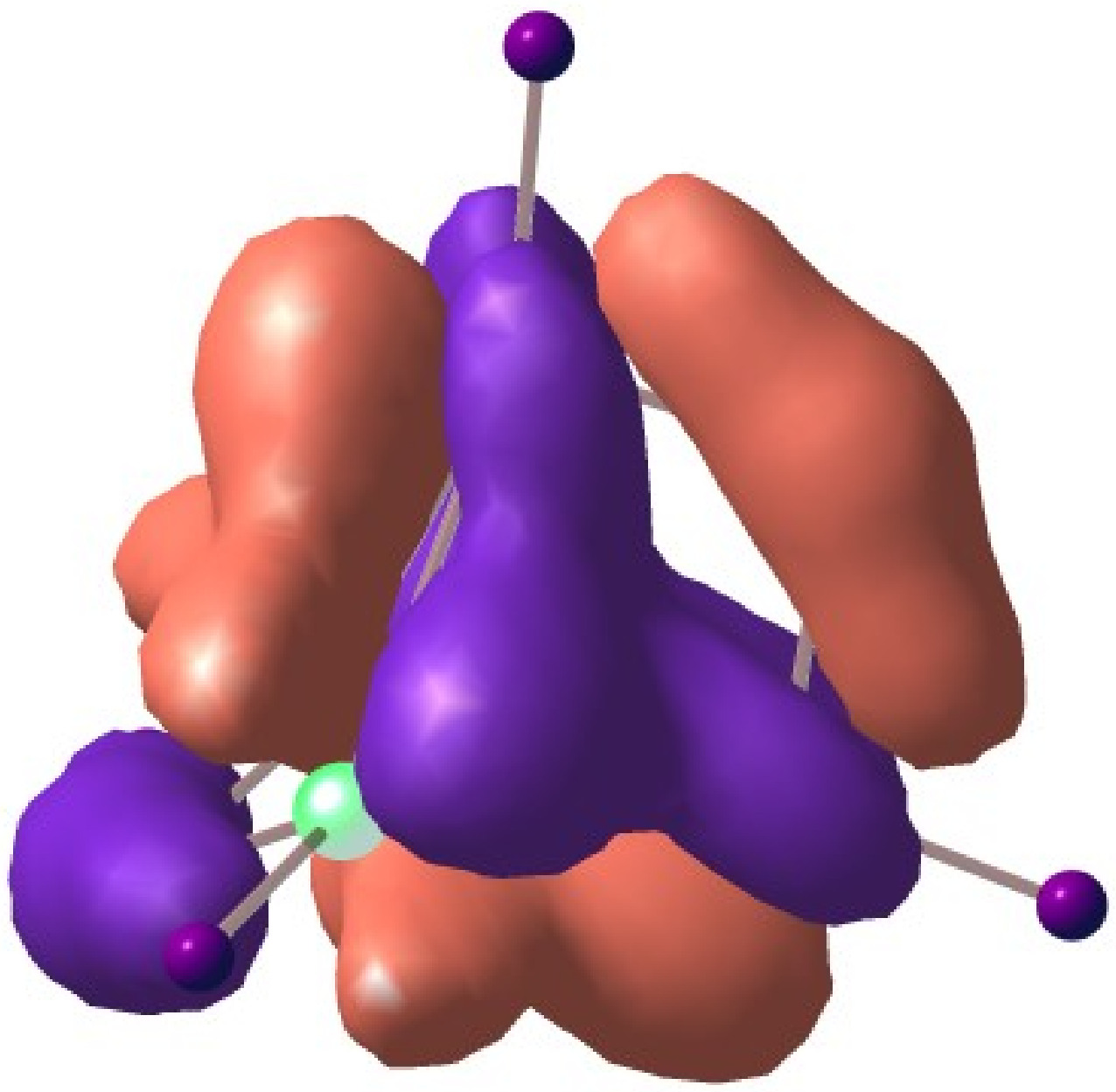}

}\subfloat[LUMO+1]{\includegraphics[width=2.25cm]{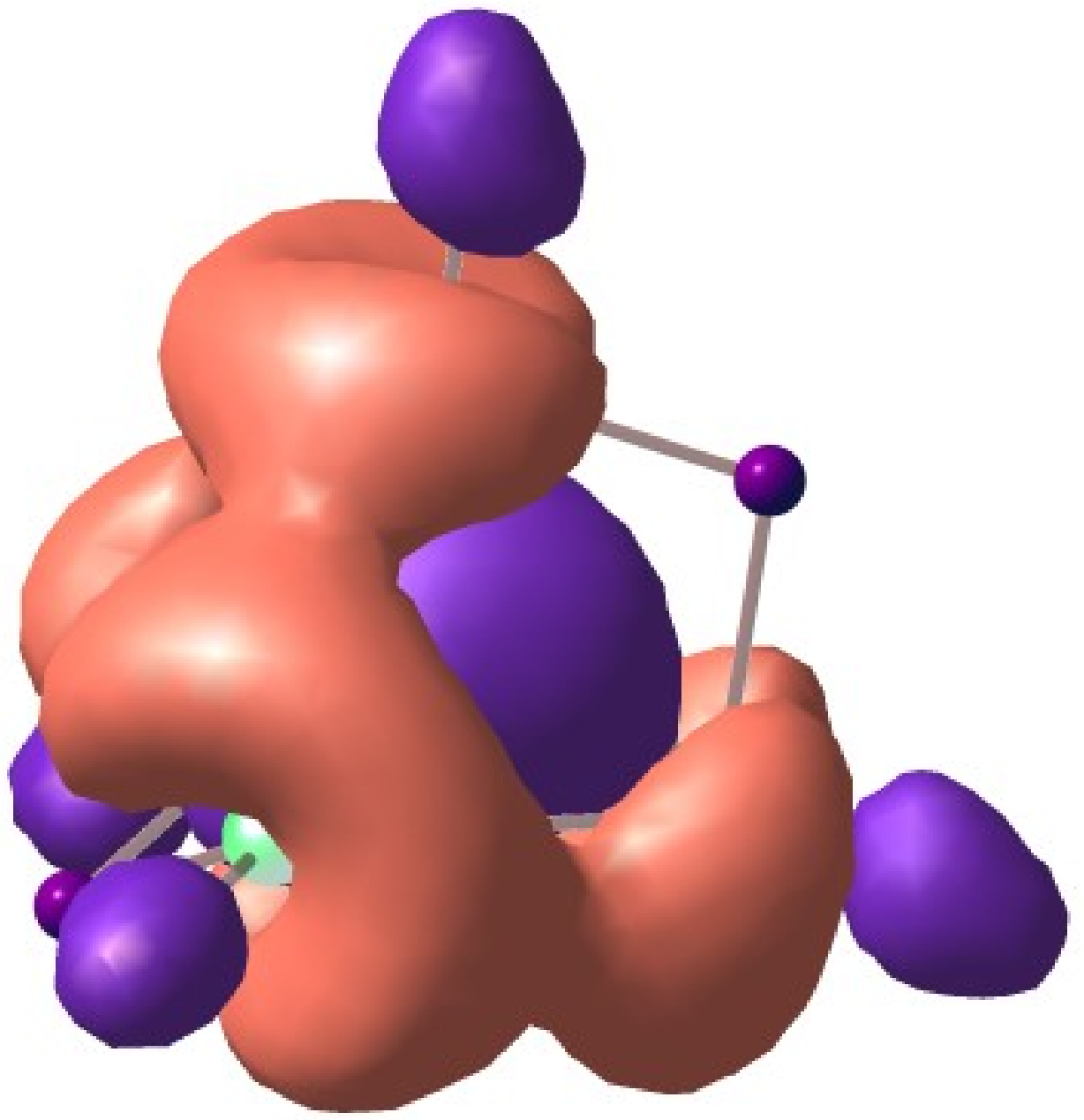}

}\subfloat[LUMO+2]{\includegraphics[width=2.25cm]{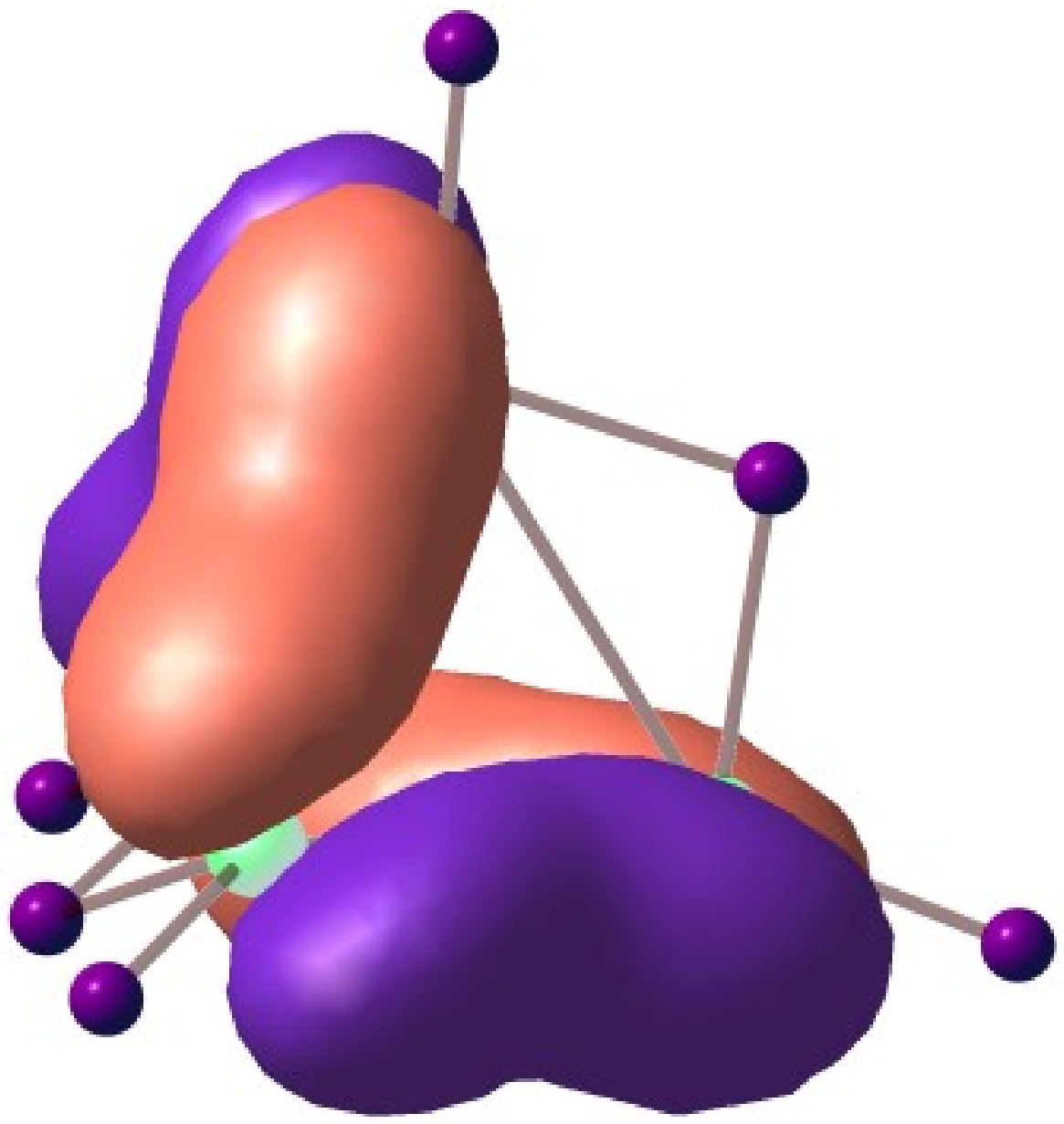}

}

\subfloat[LUMO+3]{\includegraphics[width=2.25cm]{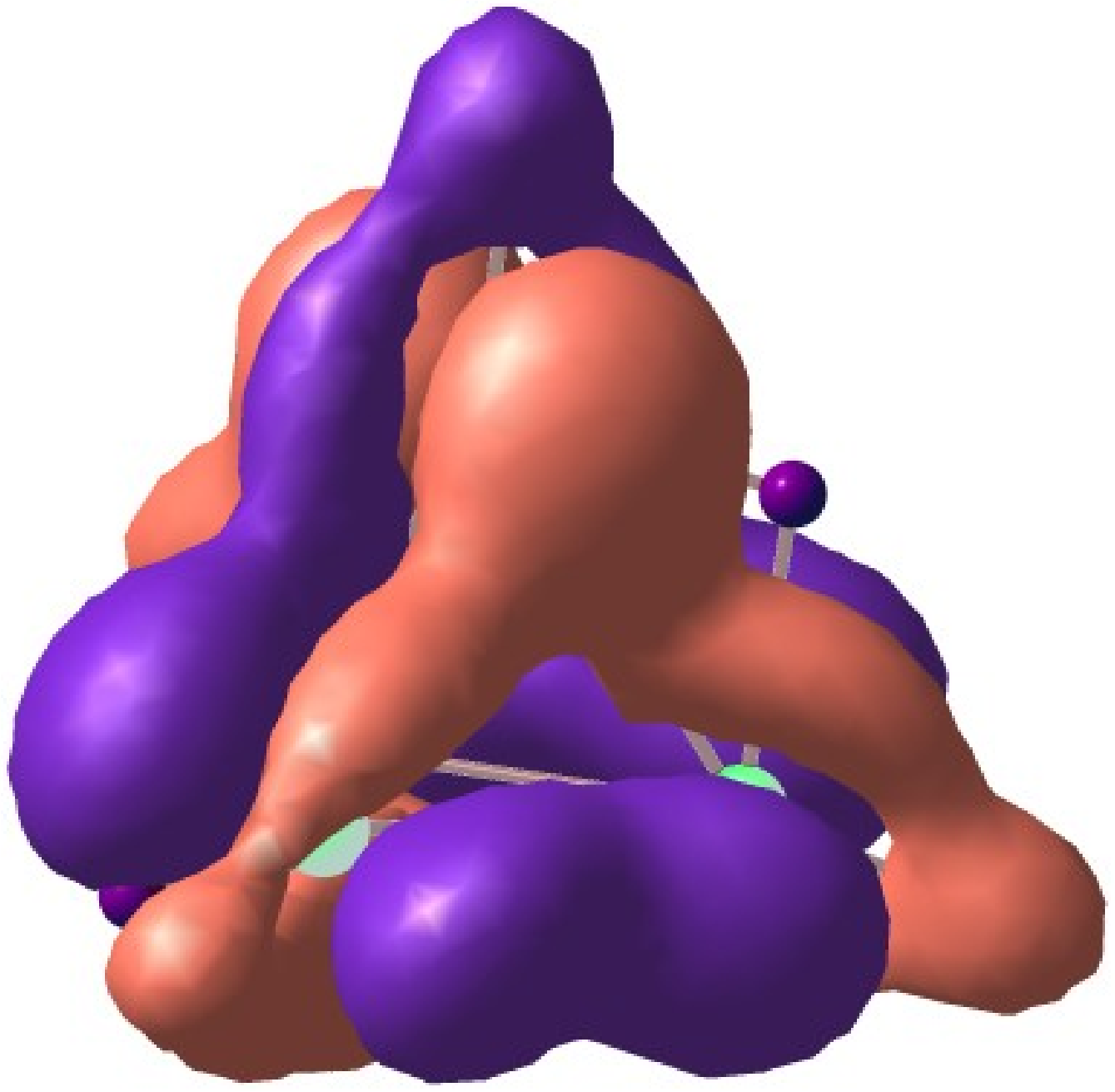}

}

\caption{(Color online) Molecular orbitals (iso plots) of Al$_{\text{4}}$H$_{\text{6}}$
from HOMO-3 to LUMO+3, obtained from the CNDO-HF calculations.}

\label{Fig:al4h6_mo_cndo}
\end{figure}

Before discussing our calculated spectra, we present some of the molecular
orbitals (MOs) of the system close to the Fermi level. In Fig. \ref{Fig:al4h6_mo_cndo}
MOs obtained from the CNDO-HF calculations are presented. Although,
here we have not presented the \emph{ab initio} B3LYP MOs which were
used to perform the TDDFT calculations, we note that among the occupied
orbitals, $H-3$, $H-1$, and $H$ obtained from the CNDO-HF and the
DFT calculations were qualitatively very similar. Mulliken charges
of various atoms indicate Al to be in slightly cationic state, while
the H atoms carry small negative charges. Therefore, we conclude that
the bonding in the system is largely covalent, with some polar character.
The covalent nature of the bonding is also obvious from the MO plots,
which show that there is significant amount of charge density in the
region between the atoms.

\begin{figure}
\vspace{1.5cm}

\includegraphics[width=8cm]{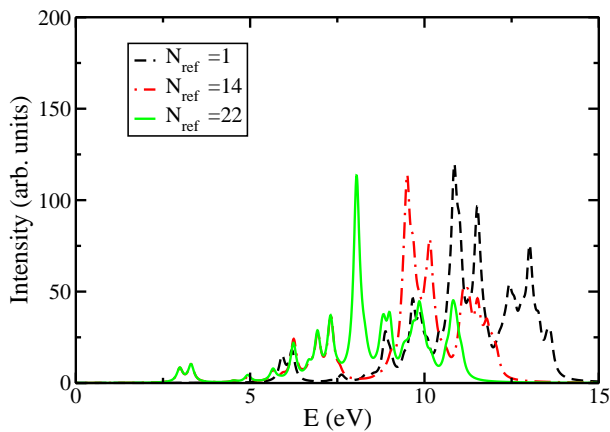}

\caption{(Color online) Convergence of the linear absorption spectrum of Al$_{\text{4}}$H$_{\text{6}}$
computed using the CNDO-MRSDCI method with respect to the number of
reference states used in MRSDCI calculations. A line width of 0.1
eV was used to compute the spectra.}

\label{Fig-al4h6_converg}
\end{figure}

Next, we present and discuss the linear optical absorption spectrum
of the system computed using the CNDO-CI and the \emph{ab initio}
TDDFT methods. The calculations were performed using the geometry
shown in Fig. \ref{Fig-struct-al4h6}, which as mentioned above, was
obtained at the B3LYP/6-311+g(d) level of theory. For the CNDO-CI
calculations the total number of valence electrons was 18 and the
total number of basis functions was 42, thus ruling out the FCI calculations.
Therefore, as mentioned in section \ref{sec:Theory-al4h6}, the MRSDCI
method was adopted to compute correlated wave functions and energies
of the ground and the optically active excited states. The calculations
were performed without the use of point-group symmetry, therefore,
the CI expansion became fairly large scale, with the largest CI calculation
consisting of 1.3 millions CSFs. The CI matrix was iteratively diagonalized\cite{meld}
to obtain 60 lowest roots which required several hours of CPU time.
The dipole moments connecting these excited states to the ground state
were used to compute the photoabsorption spectrum presented here.
Before discussing our final spectrum, in Fig. \ref{Fig-al4h6_converg}
we present the convergence of our optical absorption results with
respect to the size of the CI matrix, represented in this case by
the number of reference configurations (N$_{\text{ref}}$) from which
singly- and doubly- excited CSFs were generated to obtain the MRSDCI
expansion. For example, in our largest MRSDCI calculation we used
N$_{\text{ref}}$=22 which generated 1.3 million total CSFs. The convergence
of our calculations with respect to N$_{\text{ref}}$ is obvious from
Fig. \ref{Fig-al4h6_converg}, in which the spectra corresponding
to N$_{\text{ref}}$=14 and N$_{\text{ref}}$=22 are virtually indistinguishable. 

\begin{figure}
\vspace{0.5cm}

\includegraphics[width=8cm]{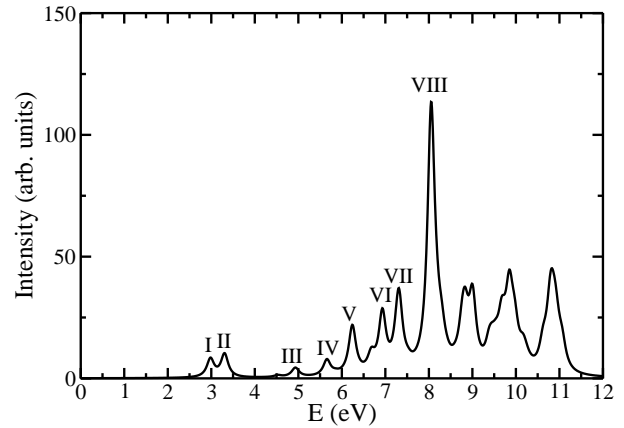}

\caption{Linear optical absorption spectrum of Al$_{\text{4}}$H$_{\text{6}}$,
computed using the CNDO-MRSDCI approach. Peaks with energies up to
8 eV have been labeled. A line width of 0.1 eV was used to compute
the spectrum.}

\label{Fig-al4h6_opt}
\end{figure}

In Fig. \ref{Fig-al4h6_opt}, we present our final linear optical
absorption spectrum of Al$_{\text{4}}$H$_{\text{6}}$ computed using
the CNDO-CI method. Before discussing our spectrum in detail, we would
like to benchmark it against other calculations to ensure its correctness.
Therefore, to ascertain the accuracy of our CNDO-CI calculations,
we also performed \emph{ab initio} TDDFT calculation of its lowest
photoexcited state, at the B3LYP/6-311+g(d) level of theory using
the Gaussian03 program package\cite{gaussian03}. Our CNDO-CI value
of 2.98 eV (peak I in Fig.\ref{Fig-al4h6_opt}), compares excellently
with the TDDFT value of 3.03 eV, which gives us confidence as to correctness
of our calculation. Martinez and Alonso\cite{AlnHn+2-martinez} used
an \emph{ab initio} TDDFT methodology to compute the photoabsorption
spectrum of Al$_{\text{4}}$H$_{\text{6}}$ all the way up to 12 eV
(\emph{cf.} Fig. 7 of Ref.\cite{AlnHn+2-martinez}). The first peak
in the TDDFT spectrum of Martinez and Alonso\cite{AlnHn+2-martinez}
is at an energy slightly higher than 3 eV, and their highest peak
occurs at an energy close to 9 eV, thus making their spectrum slightly
blue-shifted compared to ours. However, the qualitative nature of
their spectrum\cite{AlnHn+2-martinez} is quite similar to our spectrum
discussed below.

Next, we examine the nature of excited states corresponding to the
peaks present in our calculated spectrum (\emph{cf.} Fig. \ref{Fig-al4h6_opt}).
The important peaks in the spectrum up to the excitation energy $\thickapprox$8
eV have been labeled, and the many-particle wave functions of these
excited states, along with their ground-state transition dipole moments,
are presented in table \ref{tab-wave-al4h6}. The first peak in Fig.
\ref{Fig-al4h6_opt} occurs at 2.98 eV, and is a relatively weak peak
corresponding to the across the gap $H\rightarrow L$ transition.
The next peak (II) located at 3.28 eV is also similarly weak and corresponds
to $H-1\rightarrow L$ excitation. When we examine the nature of the
many-particle wave functions (\emph{cf}. Table \ref{tab-wave-al4h6})
of various excited states, we realize that: (a) the wave functions
of all the states consist predominantly of singly-exited CSFs, (b)
all the states have one dominant configuration with magnitude of its
coefficient always greater than 0.8, with some configuration mixing
with the increasing excitation energy. The highest peak in the spectrum
is located at 8.05 eV with the dominant CSF being $\mid H-1\rightarrow L+7\rangle$,
with some contributions from $\mid H\rightarrow L+9\rangle$ and $\mid H-2\rightarrow L+3\rangle$.
Koutecký and coworkers\cite{koutecky} have formulated a criterion
according to which if the many particle wave function of a given excited
state is dominated by one singly-excited configuration, it is classified
as a normal inter-band absorption. On the other hand, if the wave
function exhibits strong mixing of several CSFs with coefficients
of almost equal magnitude, it is considered to be a plasmonic collective
excitation\cite{koutecky}. In metallic clusters such as Al$_{\text{4}}$H$_{\text{6}}$,
to know whether the excited states are plasmonic in nature or not,
is always of interest. As per the Koutecký \cite{koutecky} criterion,
the excited states corresponding to peaks I to VIII do not appear
to be of plasmonic type. This is to be contrasted with the nature
of optical absorption in the bare Al$_{\text{4}}$ rhombus cluster
which were found to be of plasmonic type by Deshpande \emph{et al}.\cite{al-despandey}
in an \emph{ab initio} time-dependent local-density approximation
(TDLDA) calculation. In their work, authors reported that low-intensity
optical absorption in Al$_{\text{4}}$ begins close to 1 eV, and increases
in intensity with energy peaking around 8 eV. This suggests that in
Al$_{\text{4}}$H$_{\text{6}}$, it is the influence of hydrogenation
which increases the optical gap, as well as changes the nature of
excited states from plasmonic to inter-band excitations.

\begin{table}
\caption{Excitation energies and many-particle wave functions of excited states
corresponding to the peaks in the CNDO-CI linear absorption spectrum
of Al$_{\text{4}}$H$_{\text{6}}$ (\emph{cf.} Fig. \ref{Fig-al4h6_opt}),
along with the squares of their dipole coupling ($\mu^{2}=\sum_{i}|\langle f|d_{i}|G\rangle|^{2}$)
to the ground state. $|f\rangle$ denotes the excited state in question,
$|G\rangle$, the ground state, and $d_{i}$ is the $i$-th Cartesian
component of the electric dipole operator. In the wave function column,
the numbers in the parentheses are the CI coefficients of a given
electronic configuration. Symbols $H$/$L$ denote HOMO/LUMO orbitals.}

\begin{tabular}{|c|c|c|c|}
\hline 
Peak & Energy (eV) & $\left\langle \mu^{2}\right\rangle $ & Wave function\tabularnewline
\hline
\hline 
I & 2.98 & 0.08 & $\mid H\rightarrow L\rangle$(0.9425)\tabularnewline
\hline 
II & 3.28 & 0.09 & $\mid H-1\rightarrow L\rangle$(0.9423)\tabularnewline
\hline 
III & 4.93 & 0.02 & $\mid H-1\rightarrow L+2\rangle$(0.9226)\tabularnewline
\hline 
 &  &  & $\mid H\rightarrow L+1\rangle$(0.1531)\tabularnewline
\hline 
IV & 5.66 & 0.03 & $\mid H\rightarrow L+3\rangle$(0.9313)\tabularnewline
\hline 
V & 6.24 & 0.07 & $\mid H-1\rightarrow L+3\rangle$(0.9027)\tabularnewline
\hline 
 &  &  & $\mid H-1\rightarrow L+4\rangle$(0.1609)\tabularnewline
\hline 
 &  &  & $\mid H\rightarrow L+4\rangle$(0.1056)\tabularnewline
\hline 
VI & 6.93 & 0.11 & $\mid H\rightarrow L+5\rangle$(0.9249)\tabularnewline
\hline 
 &  &  & $\mid H\rightarrow L+4\rangle$(0.1242)\tabularnewline
\hline 
VII & 7.31 & 0.12 & $\mid H-3\rightarrow L\rangle$(0.8391)\tabularnewline
\hline 
 &  &  & $\mid H-1\rightarrow L+5\rangle$(0.3062)\tabularnewline
\hline 
 &  &  & $\mid H-2\rightarrow L+2\rangle$(0.1600)\tabularnewline
\hline 
VIII & 8.05 & 0.44 & $\mid H-1\rightarrow L+7\rangle$(0.8590)\tabularnewline
\hline 
 &  &  & $\mid H\rightarrow L+9\rangle$(0.2880)\tabularnewline
\hline 
 &  &  & $\mid H-2\rightarrow L+3\rangle$(0.1812)\tabularnewline
\hline
\end{tabular}

\label{tab-wave-al4h6}
\end{table}

\section{Conclusions\label{sec:Conclusions-al4h6}}

In conclusion, we have preseneted the linear optical absorption spectrum
of Al$_{\text{4}}$H$_{\text{6}}$ calculated using a wave function
based methodology employing the CNDO-CI approach, and compared it
with that computed using the \emph{ab initio} TDDFT approach. The
photoabsorption spectra computed using the two approaches are generally
in good agreement with each other, with the published TDDFT spectrum\cite{AlnHn+2-martinez}
being slightly blue-shifted compared to our work. Therefore, it will
be of considerable interest to perform optical absorption experiments
on Al$_{\text{4}}$H$_{\text{6}}$ cluster, to determine the nature
of its optical excitations, and shed further light on its electronic
structure. Our results can then be used for optical characterization
of this cluster.

\end{document}